\documentstyle[twocolumn,aps]{revtex}
\begin{document}
\draft
\title{\bf {Hysteresis in mesoscopic superconducting disks: the
Bean-Livingston barrier}}
\author{P. Singha Deo\cite{eml}, V. A. Schweigert \cite{sc}, and F.
M.  Peeters \cite{fmp}}
\address{Department of Physics, University of Antwerp (UIA),
B-2610 Antwerpen, Belgium.}
\maketitle
\begin{abstract}
The magnetization behavior of mesoscopic superconducting disks can
show hysteretic behavior which we explain by using the
Ginzburg-Landau (GL) theory and properly taking into account the
de-magnetization effects due to geometrical form factors. In large
disks the Bean-Livingston surface barrier is responsible for the
hysteresis.  While in small disks a volume barrier is responsible for
this hysteresis.  It is shown that although the sample magnetization
is diamagnetic (negative), the measured magnetization can be positive
at certain fields as observed experimentally, which is a consequence
of the de-magnetization effects and the experimental set up.
\end{abstract}
\narrowtext

Hysteresis in the magnetization of superconductors \cite{cod,jos,bea}
is a fascinating field of fundamental research which is related to
the occurrence of meta-stability.  Here we will investigate this
phenomenon in single mesoscopic superconducting disks.  Recently,
Geim {\it et al} \cite{exp} used the Hall probe technique to study
the magnetization of single mesoscopic $Al$ disks. The investigated
disks can be classified as few fluxoid disks (FwFD) and fractional
fluxoid disks (FrFD). The FrFD are so tiny that fluxoids cannot
nucleate in them because the required magnetic field to create a
fluxoid in it exceeds the critical field of the sample.  The FwFD are
those in which a few fluxoids can nucleate (typically of the order of
10 and definitely greater than 1) before it makes a transition to the
normal state.  It was found that these disks: 1) exhibit a variety of
phase transitions (type I or type II) that are absent in macroscopic
samples; 2) show strong hysteresis behavior, and 3) in the field down
sweep can exhibit paramagnetic behavior.

In our earlier work \cite{deo1} we presented a quantitative
explanation of the magnetization of the different disks as function
of increasing external magnetic field without using any fitting
parameters. In increasing magnetic field the experimental data were
found to follow the magnetization of the ground state of the sample.
Axially symmetric solutions with a fixed angular momentum $L$ were
assumed and the non-linear GL eqns. were solved for disks with a
finite thickness. Switching between the different $L$ states occurs
when their free energies are equal leading to steps in the
magnetization of the mesoscopic superconductor.  In increasing field
the {\it Bean Livingston} (BL) surface barrier responsible for
meta-stability, is destroyed by boundary roughness which explains why
the system follows the ground state.  The BL barrier arises from the
fact that the superconducting currents around a vortex is in the
opposite direction to the screening currents at the surface of the
sample which does not allow the nucleation of vortices at the
boundary, although the free energy is lowered when the vortex moves
to the center of the sample.

The decreasing field behavior of the magnetization, which was not
studied in our previous work, can be very different and can even show
paramagnetic behavior.  One of the reasons for this different
behavior is that the Bean Livingston barrier in this case is not
destroyed by surface defects and the system will not evolve along the
ground state.  A simple estimate of the BL barrier using the approach
outlined in Ref. \cite{bea} shows that it can be several orders of
magnitude larger than $kT$ for experiments done at 0.4K.  In this
case the steady state will not necessarily be the ground state but
will be an excited state determined by the history of the sample and
the meta-stability created by the Bean Livingston barrier.

Our theoretical approach was outlined in in Refs. \cite{deo1,sch}.
The order parameter is considered to be uniform in the z direction
which is a very good approximation for thin disks, i.e., thickness
less than the coherence length.  We use the Gauss-Seidel method to
solve the non-linear GL eqn. (Eqn. (1) in Ref \cite{deo1}) and the
fast Fourier transform to solve the 3D Maxwell eqn. (Eqn. (2) in Ref.
\cite{deo1}). The order parameter of the previous magnetic field is
taken as the initial order parameter for a particular magnetic field.
This ensures that the system does not escape from the barriers
leading to meta-stability and converges differently in increasing and
decreasing fields. A large number of iterations (typically of the
order of 5x$10^4$) are then made to arrive at a self consistent
solution.

As a typical case let us consider a FwFD (radius $R$=0.8$\mu m$,
thickness $d$=0.134 $\mu m$, coherence length $\xi(0)$=0.183 $\mu m$
and penetration length $\lambda(0)$=0.07$\mu m$) with parameters
comparable to one of the disks used in the experiment \cite{exp}. The
solution assuming axial symmetry is referred to as the 2D solution
(due to the symmetry the azimuthal direction drops out) whereas the
general solution without assuming axial symmetry will be referred to
as 3D solution.  Let us consider the 2D solution first. The
dimensionless free energy G, in units of $H_c^2V/8\pi$ \cite{deo1}
(here V is the volume of the sample), is shown in Fig. 1(b) by the
thin solid curves as a function of the applied magnetic field for the
different $L$ states.  The corresponding magnetization for these
different $L$ states is shown in Fig.~1(a) by the thin solid curves.
Hence from Fig.~1(b), it can be seen that up to a magnetic field of
42.6 Gauss, the $L$=0 state is the ground state.  Beyond this field
the $L$=1 state becomes the ground state. As we increase the field,
the higher $L$ states become the lowest energy state. This continues
as long as the free energy G is negative. A positive G means that the
normal state has a lower free energy than the superconducting state
and the superconductor turns normal. The free energy of the ground
state of the system is therefore given by the thick solid curve in
Fig.~1(b).  The corresponding magnetization along this ground state
is given by the thick solid curve in Fig.~1(a). However the free
energy and the magnetization in increasing magnetic field as given by
the 3D solution is given by the thick dotted curve in Fig.~1. Thus
the 3D solution in increasing magnetic field, takes the system along
a steady state that conserves $L$ up to the point where the free
energy is zero at which point a jump to a higher $L$ state occurs.
The free energy and magnetization curve in decreasing magnetic field
as given by the 3D solution is given in Fig. 1 by the thick dashed
curves.  It can be seen that along the steady state the system moves
along states whose free energy is much higher than that of the ground
state.

The experimental data for the disk considered in Fig. 1, in
decreasing and increasing field are given in Fig. 2 by open circles
and squares, respectively. These curves are plotted according to the
scale on the left y-axis. The dashed and dotted curves in Fig.~2 are
the same as the thick solid and thick dashed curves in Fig. 1(a),
their scale being depicted on the right y-axis.  The magnitude of the
measured magnetization (open circles and squares) is clearly too
small compared to the sample magnetization in increasing and
decreasing fields (dashed and dotted curves).  Furthermore in the
experimental data, paramagnetic behavior is found for certain
magnetic fields.  Although diamagnetism is a fundamental property of
superconductors, previously a paramagnetic Meissner effect \cite{tho}
was observed on large Nb disks. This discovery lead to intensive
research but the effect is still not completely understood
\cite{mos1}. In the presence of pinning also superconducting samples
can exhibit paramagnetic behavior. But we found that this discrepancy
(i.e., the factor of 25 and the paramagnetic behavior) can be
explained by considering the full experimental set up of Ref.
\cite{exp}.

The magnetometry used in the experimental work of Ref \cite{exp} is
explained in detail in Ref. \cite{mag}. The superconducting sample is
mounted on top of a small ballistic Hall cross and the magnetization
of the superconducting disk was measured through the Hall effect. In
Ref. \cite{lee} it was shown that the Hall voltage of a Hall cross,
in the ballistic regime, is determined by the average magnetic field
piercing through the Hall cross region.  The Hall cross has a larger
area than the sample and it measures the magnetization of this area
rather than the magnetization of the sample. The effective area of
the Hall cross is approximately 3 $\mu m$x3 $\mu m$. The field
distribution in case of thin disks is extremely non-uniform inside as
well as outside the disks and the detector size will have an effect
on the measured magnitude of the magnetization, the nature and extent
will depend on the magnetic field profile outside the sample.  We
calculate the magnetization measured by the detector by integrating
the field expelled from its area, i.e., the Hall cross.  We took the
detector to be a square whose side is of length 3.1 $\mu m$ and is
placed just below the superconducting disk.  The resulting
magnetization is given by the solid curves in Fig. 2 drawn according
to the scale on the left y-axis i.e., with the same scale as the
experimental data.  Note that by including the effect of the
detector, the magnetization 1) is scaled down considerably, 2) the
line shape is changed slightly, and 3) the detector output can give a
positive magnetization although the sample itself is diamagnetic.
Note that these three factors bring the theoretical result very close
to the experimental result which explains even the apparent
diamagnetic behavior.  Only at the position of the last jump there is
a noticeable difference. This may be due to some pinning center near
the center of the disk that becomes effective when the giant vortex
state shrinks to the center.  We found that decreasing the detector
to an area enhances the apparent paramagnetic behavior. In the inset
of Fig. 2 we compare the magnetization for a detector size of 2$\mu
m$x2$\mu m$ (dotted curve) with that for a detector size of 3.1$\mu
m$x3.1$\mu m$ (solid curve).

In order to explain the reason why a larger detector can result in an
apparent paramagnetic behavior we show in Fig. 3 the total magnetic
field (applied field plus the field due to the magnetization of the
sample) along a line passing through the center of the disk. Note
that far away from the disk region the field is equal to the applied
field which is 59.65 Gauss and which is a field where the detector
magnetization shows a paramagnetic behavior. In the center of the
disk the magnetic field is much larger than the applied field because
of the flux trapped by the BL barrier. Hence the central region is
paramagnetic.  Near the edge of the disk, the magnetic field is much
smaller than the applied field and is thus a diamagnetic region. It
is this region where superconductivity is maximum and the disk is in
the giant vortex state. The sample magnetization is the resultant
magnetization of these two regions, and the total magnetization turns
out to be diamagnetic. But when we calculate the detector
magnetization, we have to include the magnetic field in the region
outside the disk which is strongly paramagnetic.  The magnetic field
just outside the disk is larger than the applied field because of the
strong flux expulsion from the disk and the important
de-magnetization effects in finite thickness disks. Or in other words
in this region the magnetization due to the currents flowing in the
disk is added to the external field and both are in the same
direction. The 2 $\mu m$ detector show greater paramagnetism that the
3 $\mu m$ detector due to the fact that over a very large area the
net flux expelled should be zero (magnetic field far away from the
sample decreases as $1/r^3$ and area of a detector increases as
$r^2$. Consequently magnetization will disappear as $1/r^5$).
However, for sufficiently small detectors the paramagnetic behavior
will disappear when its size becomes comparable to the sample size.
In the experiment of Thompson {\it et al} \cite{tho}, a paramagnetic
Meissner effect was observed when Nb disks were cooled down from
their normal state to the superconducting state, in the presence and
the absence of an external magnetic field.  The paramagnetic field
outside the sample due to the diamagnetic currents inside the sample
will be large for those large disks (R=3.2 mm).  Hence our
explanation may also be relevant for their experiment.

From the above discussion of the magnetic field distribution in and
around the superconducting disk one may ask the question whether one
can observe the superconductor in a state such that the paramagnetic
region at the center has a larger contribution than the diamagnetic
region at the boundary \cite{mos1}. In this case the sample itself
would be paramagnetic.  Indeed when the thickness of the sample is
greatly reduced, the sample magnetization itself can be paramagnetic
in a field down sweep.  We reduced the thickness by a factor of 10
and kept all other parameters fixed which lead to the sample
magnetization as shown in the inset of Fig. 3. The maximum
paramagnetism is 0.027 G which occurs at 13 G and which is indicated
by a small circle in the figure. For such a thin disk, the
diamagnetic response is very small and the flux trapped inside the
giant vortex state determines the sign of the response. The magnetic
field distribution for this case is given in Fig. 3 by the dotted
curve.  One can easily notice the weak flux expulsion from the
diamagnetic boundary and as a result the paramagnetic region outside
the sample is negligible.

Next we consider a FrFD (radius $R$=0.5$\mu m$, thickness $d$=0.15
$\mu m$, coherence length $\xi(0)$=0.25 $\mu m$ and penetration
length $\lambda(0)$=0.07$\mu m$) with a larger coherence length than
the previous one, which is probably due to smaller disorder in the
disk.  The disk shows (see Fig. 4) a first order phase transition to
the normal state.  Its behavior in increasing field was explained in
Ref. \cite{deo1}.  Hysteresis in case of this FrFD is novel because
the BL barrier cannot result in meta-stability here.  The BL barrier
requires the existence of a vortex (which is not present inside the
disk) and opposite flowing currents inside the sample.  In the
following we give an explanation for this hysteresis. For the FrFD we
use the 2D solution again, because of its high accuracy and because
it is correct in the absence of multiple vortices.  The experimental
data in increasing and decreasing fields are shown in Fig. 4 by the
squares and circles, respectively.  The 2D solution in increasing and
decreasing fields are given by the dashed and solid curves,
respectively, where the detector size is included in our
calculations.  We are using a square detector of width 2.9 $\mu m$
placed just below the sample. Although the position of the jump in
decreasing field does not coincide with the experiment the non-linear
line shape and magnitude of the magnetization is nicely reproduced.
A straight dotted line is drawn tangent to the experimental data at
the origin as a guide to the eye in order to accentuate this
non-linear behavior.  The origin of this hysteresis is due to
meta-stability created by a {\it volume barrier} and not the BL
surface barrier. In the present case of a first order transition, the
free energy has two local minima which correspond to two different
values of the order parameter which correspond to the normal and
superconducting states, respectively. These two minima are separated
by a maximum which acts as a barrier when the system tries to switch
from one minimum to the other at the critical point. To differentiate
it from the surface barrier we call it volume barrier. As this
barrier only appears near the critical field for each $L$ state, this
barrier is not relevant for the FwFD where transitions between
different $L$ states occur well below the critical field for them.
It is true that in decreasing field the position of the jump does not
coincide with that of the experiment but this position is extremely
dependent on small fluctuations in the normal system where the order
parameter is zero.  The order parameter has to jump from zero to a
large value which is difficult in the absence of some external
nucleation center. If the order parameter starts growing in some
region of the sample, the neighboring regions tries to destroy it
because that will reduce the gradient term in the Free energy.
Although if the order parameter at every point is large then the free
energy is reduced because of a reduction in the potential energy.
Therefore, starting from different initial conditions we obtain
different positions where this jump occurs in decreasing fields. If
we start from smaller values of the order parameter, the jump in
magnetization moves towards smaller magnetic fields and consequently
increases the hysteresis.  The position of this jump can also be
changed if we add random fluctuations of different magnitudes to the
order parameter.  The position given here corresponds to the field
when the volume barrier disappears.

In conclusion, we showed that the hysteresis observed in mesoscopic
disks \cite{exp}, can be explained by considering meta-stability
created by energy barriers: in the FwFD it is the BL surface barrier,
whereas in the FrFD it is the volume barrier.  In the ground state
the magnetization of those disks is diamagnetic.  Sweeping down the
magnetic field brings the system in meta-stable states which have a
substantial smaller diamagnetic behavior and can even be paramagnetic
for certain thin disks. The size of the detector, has a substantial
influence on the magnitude of the measured magnetization and can even
change the sign of it.

This work is supported by the Flemish Science Foundation(FWO-Vl)
grant No: G.0232.96, the European INTAS-93-1495-ext project, and the
Belgian Inter-University Attraction Poles (IUAP-VI).  One of us (PSD)
is supported by a post doctoral scholarship from the FWO-Vl and FMP
is a Research Director with the FWO-Vl. Discussions with Dr. A. Geim
is gratefully acknowledged.

\centerline{FIGURE CAPTIONS}

Fig. 1. The magnetization (a) and dimensionless free energy in units
of $H_c^2V/8\pi$ (b) for a FwFD (radius $R$=0.8$\mu m$, thickness
$d$=0.134 $\mu m$, coherence length $\xi(0)$=0.183 $\mu m$ and
penetration length $\lambda(0)$=0.07$\mu m$) as a function of
magnetic field. The thin solid curves are the 2D solutions for
different $L$ states. The ground state is given by the thick solid
curve.  The dotted and dashed curves are the increasing and
decreasing field behavior obtained from the 3D solution.

Fig. 2. The experimental results for the magnetization (left scale)
in increasing (circles) and decreasing (squares) magnetic field of
the disk of Fig. 1. Theoretical results for the sample magnetization
(right scale) are given by the dashed and dotted curves. Inclusion of
the detector size, gives the corresponding thin and thick solid
curves with reference to the left scale.

Fig. 3. The solid curve is the field distribution (applied field of
59.65 G plus magnetic field due to the magnetization of the sample)
along a line through the center of the disk of Fig. 1. The disk
region and the detector region are indicated. The dotted curve is the
corresponding curve at an applied field of 13 G for a disk which is
ten times thinner. The inset shows the magnetization versus
decreasing applied field for this thin sample.

Fig. 4. The squares (circles) are the experimental data in increasing
(decreasing) field for a FrFD (radius $R$=0.5$\mu m$, thickness
$d$=0.15 $\mu m$, coherence length $\xi(0)$=0.25 $\mu m$ and
penetration length $\lambda(0)$=0.07$\mu m$).  The dashed (solid)
curve is the corresponding numerically calculated detector
magnetization.

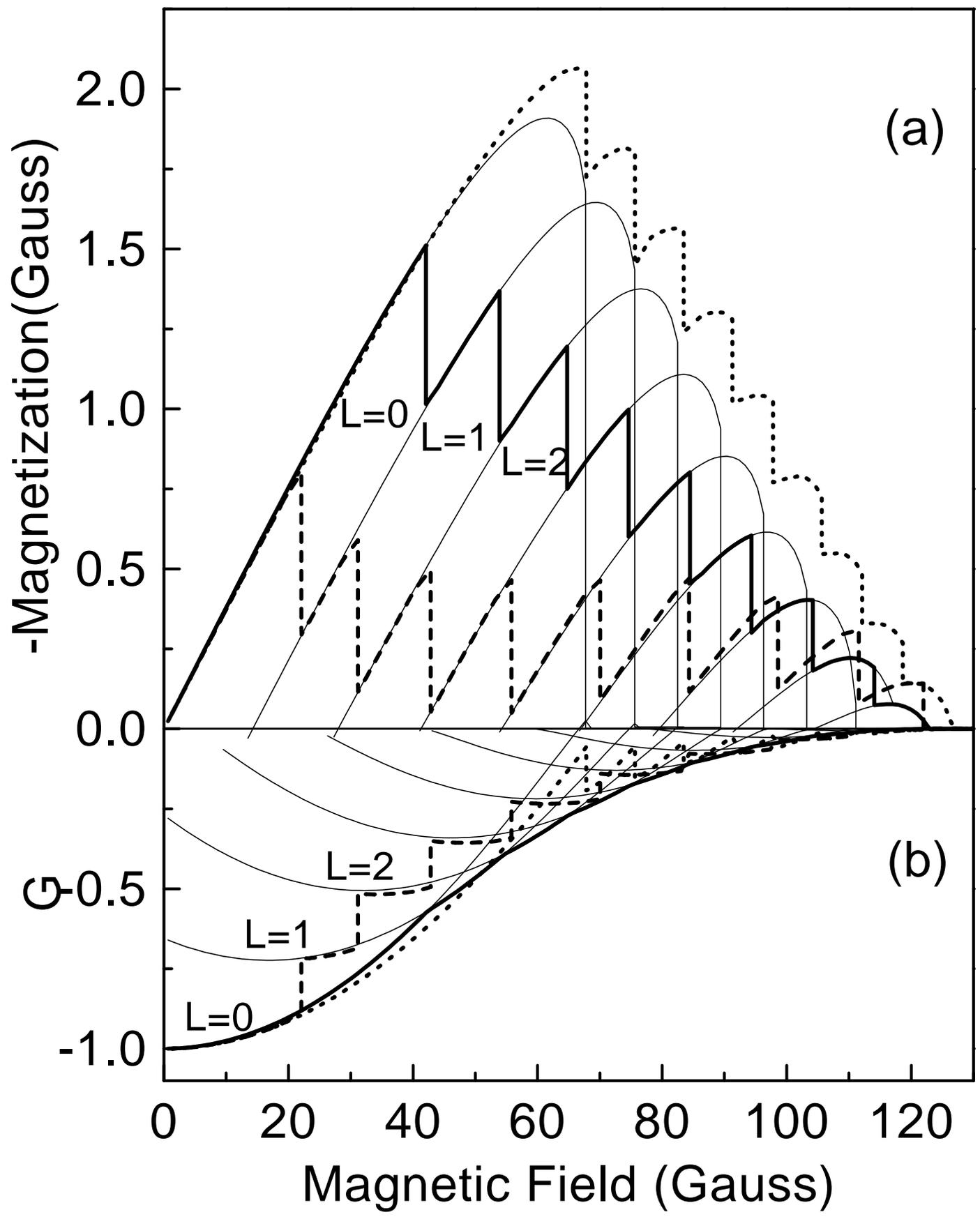
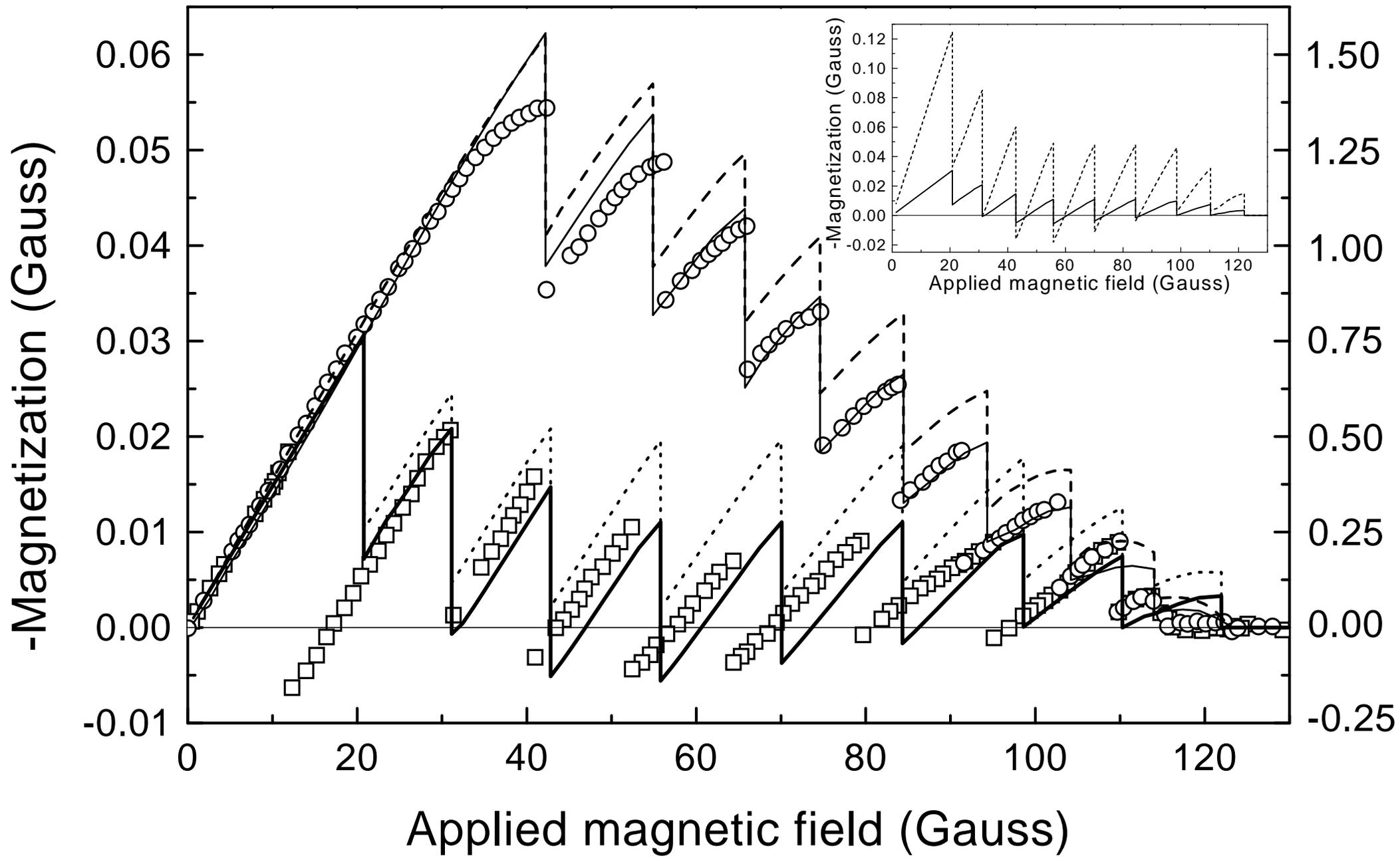
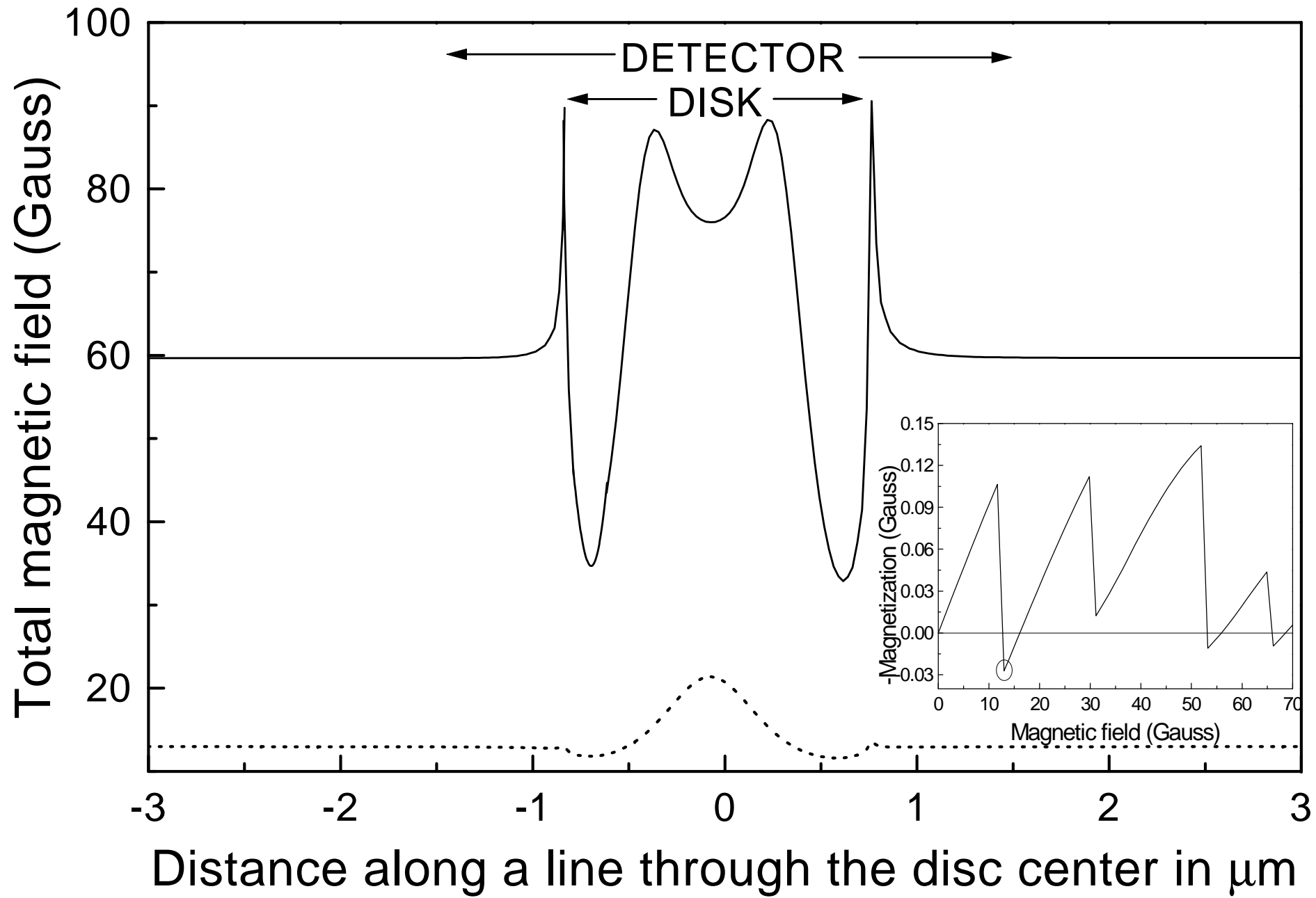
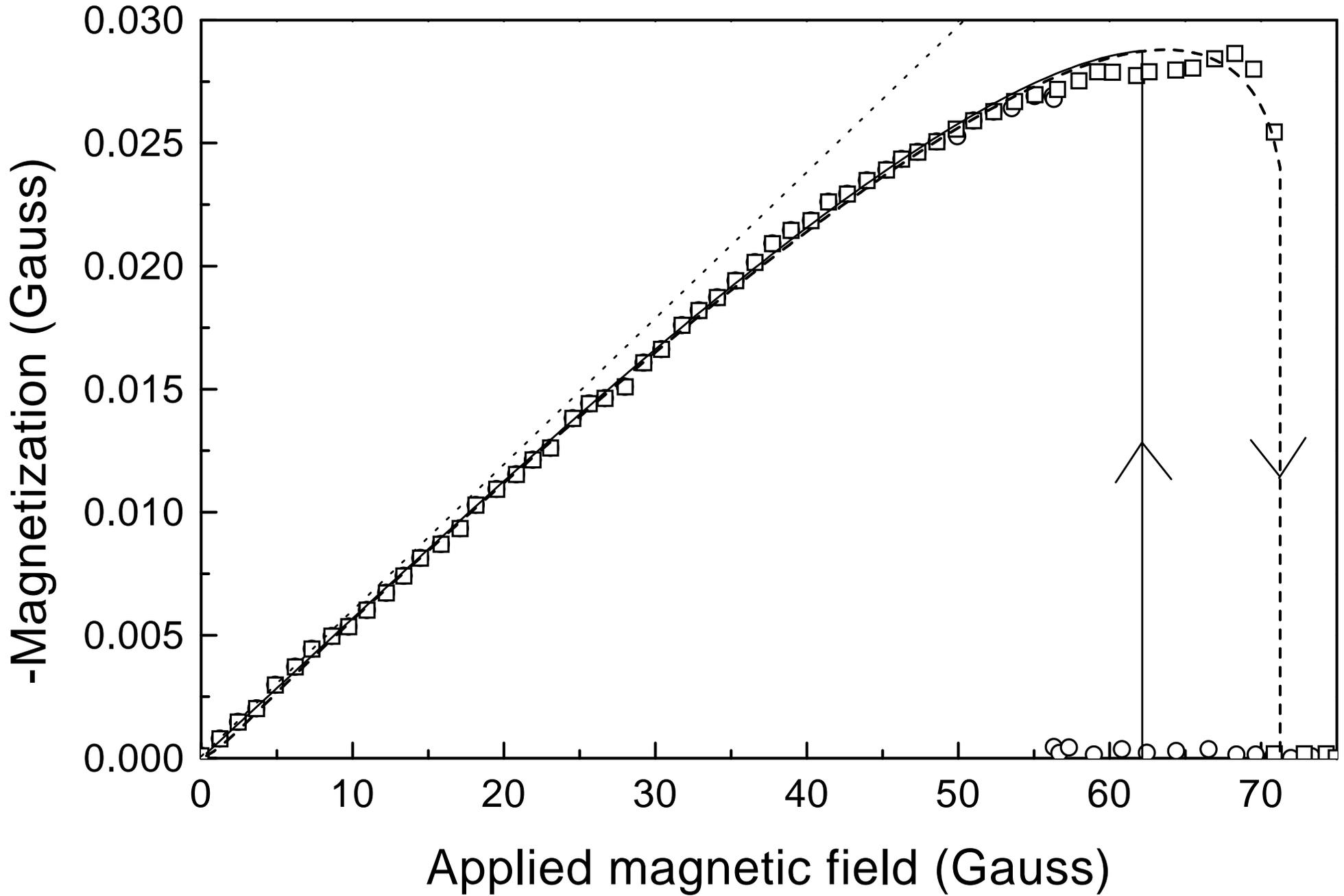
\end{document}